# Spiral arm structures revealed in the M31 galaxy

## Yu.N.Efremov

*Sternberg Astronomical Institute, MSU, Universitetsky pr. 13, Moscow, 11992 Russia*

## Abstract

Striking regularities are found in the northwestern arm of the M31 galaxy. Star complexes located in this arm are spaced 1.2 kpc apart and have similar sizes of about 0.6 kpc. This pattern is observed within the arm region where Beck et al. (1989) detected a strong regular magnetic field with the wavelength twice as large as the spacing between the complexes. Moreover, complexes are located mostly at the extremes of the wavy magnetic field. In this arm, groups of HII regions lie inside star complexes, which, in turn, are located inside the gas–dust lane. In contrast, the southwestern arm of M31 splits into a gas–dust lane upstream and a dense stellar arm downstream, with HII regions located mostly along the boundary between these components of the arm. The density of high-luminosity stars in the southwestern arm is much higher than in the northwestern arm, and the former is not fragmented into star complexes. Furthermore, signatures of the age gradient across the southwestern arm have been found in earlier observations. This drastic difference in the structure of the segments of the same arm (Baade's arm S4) is probably due mostly to their different pitch angles: the pitch angle is of about $0°$ for northwestern part of the arm and about $30°$ in the southwestern segment. According to the classical SDW theory, this might result in lower SFR in the former and in the triggering of high SFR in the latter. Data on M31, M51, M74, and some other galaxies suggest that star complexes are mostly located in the arm segments that are not accompanied by a dust lane upstream – i.e., those that do not host spiral shock wave. The same is surely related to the tidal arms which often host a chain of complexes. The regularities in the distribution of complexes along a density wave spiral arm are most probably due to the development of the Parker-Jeans instability, which builds up star complexes if the initial SFR is low and, as a sequence, magnetic field is regular along the arm.

# 1 Introduction

Walter Baade (1963), who was the first to establish the concentration of HII regions along the spiral arms of the Andromeda galaxy, noted that a spiral arm behaves similarly to a chameleon: in some segments it is filled with stars, in others it finds out only on presence of the dust clouds. He meant mostly the arm which he designated as S4. We shall try to suggest the possible reasons of inconstancy of this arm structure along its length.

The HII regions and OB stars are indeed known to be distributed along the spiral arms of galaxies nonuniformly; they often form groups with sizes of about 0.5–1.0 kpc; a bit older objects, such as Cepheids, are usually also concentrated in the same groups - star complexes (Efremov 1978, 1979). These complexes - the greatest coherent groupings of stars, connected by unity of an origin from the same $HI/H_2$ supercloud (Efremov 1989, 1995; Odekon 2008, Elmegreen 2009). Sometimes the complexes located within an arm at semi-regular distances. This is a rather rare phenomenon; Elmegreen and Elmegreen (1983) found such chains of complexes in 22 grand design galaxies – out of some 200 (BGE, private communication) suitable galaxies studied in the Palomar Sky Survey photos. They found the spacing of complexes in studied galaxies to be within 1 – 4 kpc. The gravitational or magneto-gravitational instability developing along the arm was suggested to explain this regularity (Elmegreen 1994).

Star complexes in the M31 galaxy were first identified by Sidney van den Bergh (1964) who called them OB associations. However most of his groupings should be classified by all their characteristics, including their sizes, as star complexes; bona fide O- associations (with average sizes of about 80 pc) are concentrated within the complexes (Efremov et al., 1987).

The issue of whether or not there is the continuous sequence of star groups with increasing age (of the oldest stars) and size, starting from clusters to associations to complexes, is somewhat debatable (Efremov and Elmegreen, 1998); moreover, in the flocculent galaxies the largest complexes transform to short spiral segments (Elmegreen and Efremov, 1996). At any rate, within the spiral arms of the grand design galaxies, the star complexes are clearly the largest and well determined groupings of the young stars. The connection between star complexes and super associations is discussed elsewhere (Efremov 2004).

Star complexes, first picked out in the local Galaxy from distribution of Cepheids and partly of star clusters and supergiants (Efremov, 1978) were surely found within the local segments of the spiral arms, but about nothing may be said on their locations in Galaxy at large distances. This is not the case for the HI superclouds, which were first detected in the outer arms of the Milky Way galaxy by McGee and Milton (1964). The mean mass of these superclouds was found to be $10^7$ solar masses. Later on the chains of HI superclouds have appeared in Grabelsky's et al. (1987, 1988) images of HI distribution within the Carina arm. The regular spacing of these superclouds was studied by Efremov (1998), and later on – in the Cygnus (Outer) arm (Efremov 2009); the values of supercloud spacings were found to have the curious bi-modal distribution.

Here, we will provide more data on the complex locations along the spiral arms of the Andromeda galaxy. The regularities in distributions of complexes (and their very presence) in M31 are found to be connected with the local values of the arm pitch-angle and magnetic field, as well with the position of dust/gas lane in an arm segment cross-section. Similar correspondences are noted in a few other galaxies. Some preliminary results of this study were published (Efremov 2009).

## 2 Star complexes in the northwestern arm of M31

There is one more, quite simple explanation why the chains of complexes are rarely observed, as well as chains of HI superclouds. The common optical images of M31 display rather only one complex (NGC 206) which is, better to say, is not a complex but the long segment of the arm with high density of the luminous stars (Fig1a). However, in images which fixed only rather young (up to ~100 Myr) stars the complexes are evident and numerous (Figs. 2, 3). This makes evident that complexes are concentrations of the younger stars only. The HII regions mostly forms groups which are always coincide with a star complex. An individual HII region corresponds to an OB-association and this make clear that a complex may be called a group of associations, it is the highest level in the hierarchy of stellar groupings. We would like to stress that the detection of complexes in UV images only makes evident that complexes are concentrations of the younger stars only.

**a**                                        **b**

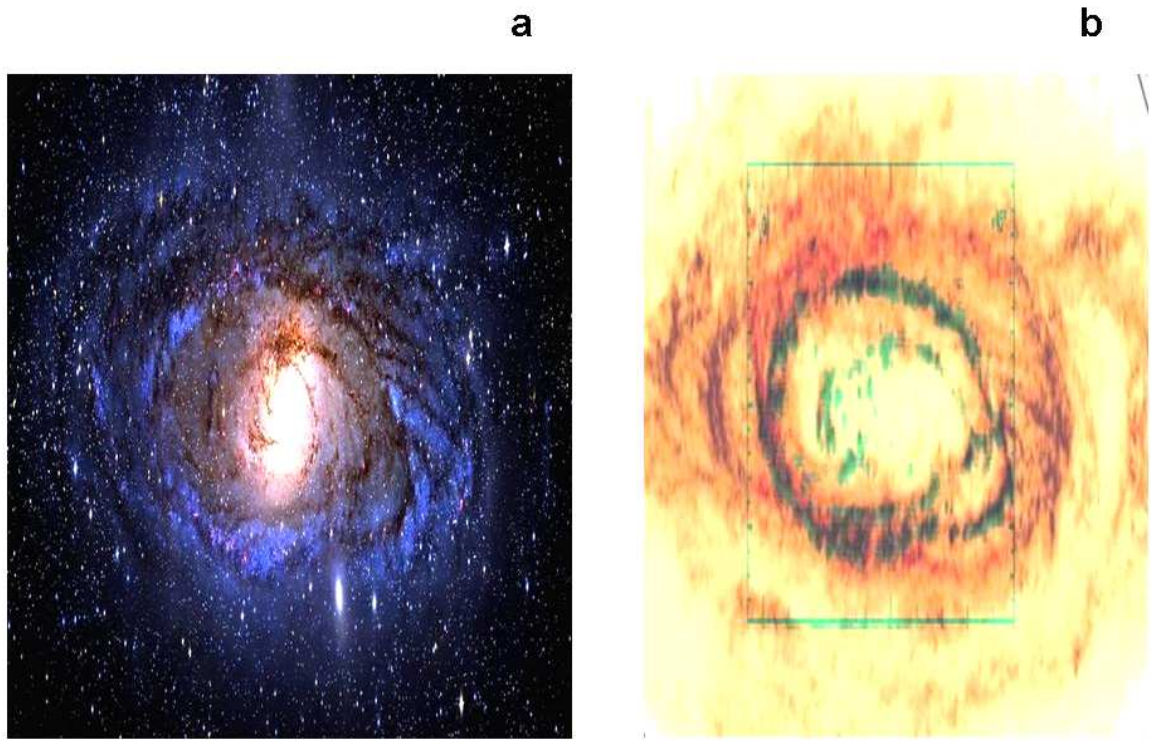

**Figure 1**. Pole-on  images of the M31 galaxy.  The inclination angle is assumed to be 12 degrees. The  NW  is at the top, and the SW  is to the right-hand side.
  (a) optical image (transformed from the image, obtained by Tony Hallas, with his kind permission).
  (b)  CO   image (transformed from Neininger et al. 2000, green) overlaid onto HI   image (transformed from Braun et al. 2009, brown).  The dust lanes closely correspond to gaseous ones, providing the stellar background is bright enough.

     Figs. 2 represents the   northwestern  part of M31 in different wavelengths.  The isolated star complexes are well seen  in  the Swift and  GALEX telescopes far-UV images and more so, in   overlay of the GALEX and Spitzer (far-IR) images, taken from the Spitzer telescope site.  The regular spacing of the complexes  along this segment of Baade's  S4 arm is now    quite evident  (Efremov, 2009).    In the right part of  this arm in Fig. 2e concentration of  HII regions inside the complexes is evident too.
     It looks now quite surprising why this regularity  was not noted early, in spite it is well seen  in  GALEX images obtained and published rather long ago.  Moreover, as Fig. 3 demonstrates,  the complexes seen in Fig.  2 and 3  are  closely matched to those outlined by Efremov et al. (1987) by eyes,  using  the large-scale  UBV plates of 2-m reflector of Rozhen Observatory (Bulgaria).  Also, most of our  complexes  correspond well to van den Bergh's associations or their  groupings, and we have preserved van den Bergh's (1964) designations. We stress  that  the regular spacing of complexes is  seen  in one arm segment only.

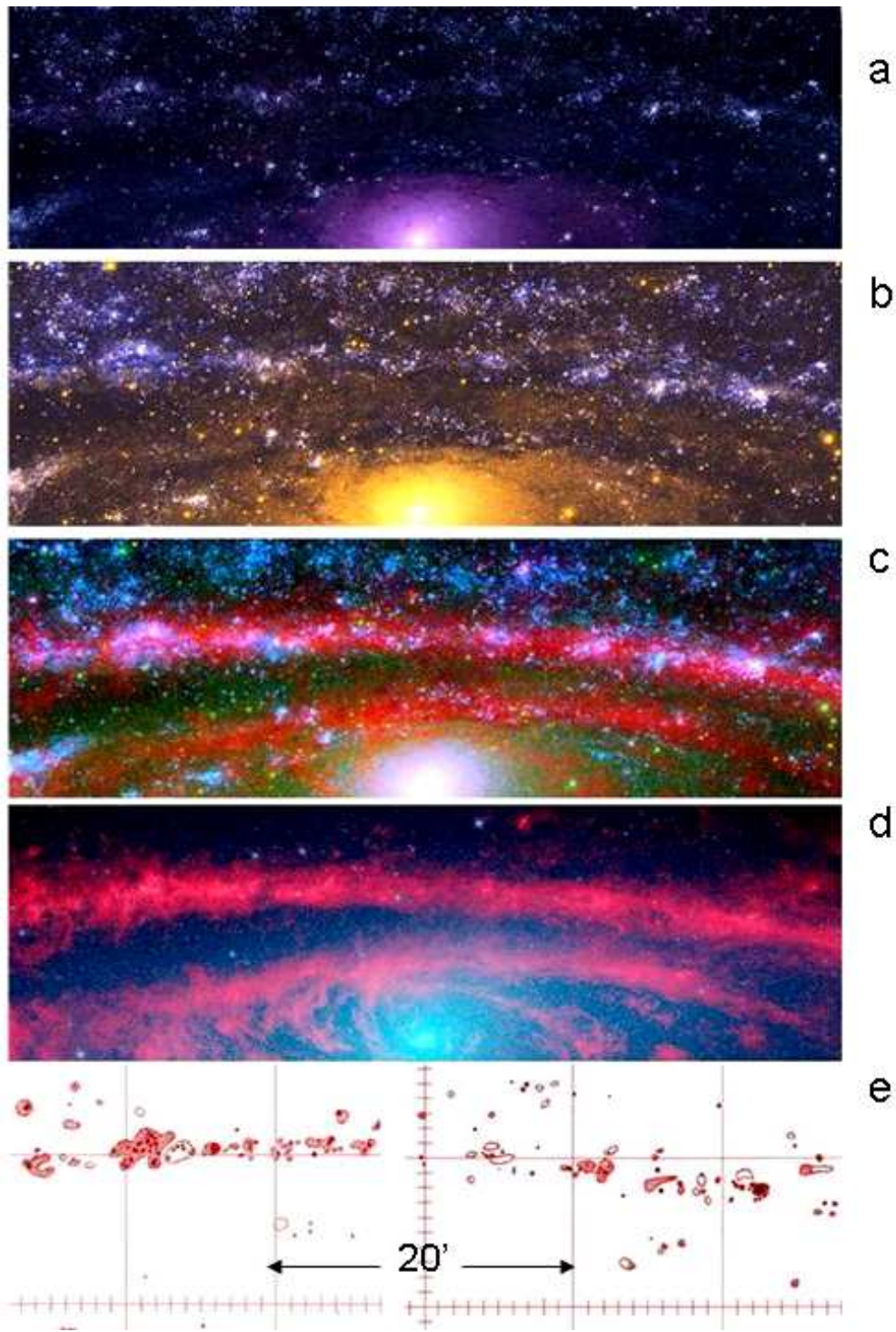

**Figure 2.** Images of the central NW segment of M 31, reduced to the same scale:
(a) The recent Swift telescope image, far UV
(b) GALEX, far-UV image (from the GALEX telescope site)
(c) GALEX overlaid onto Spitzer image (from the Spitzer telescope site).
(d) Spitzer telescope far-IR image (from the Spitzer telescope site). Note holes in
the warm dust, corresponding to brighter regions in Fig. 2c and to HII regions in Fig. 2e.
(e) HII regions (from Pellet et al. 1978).

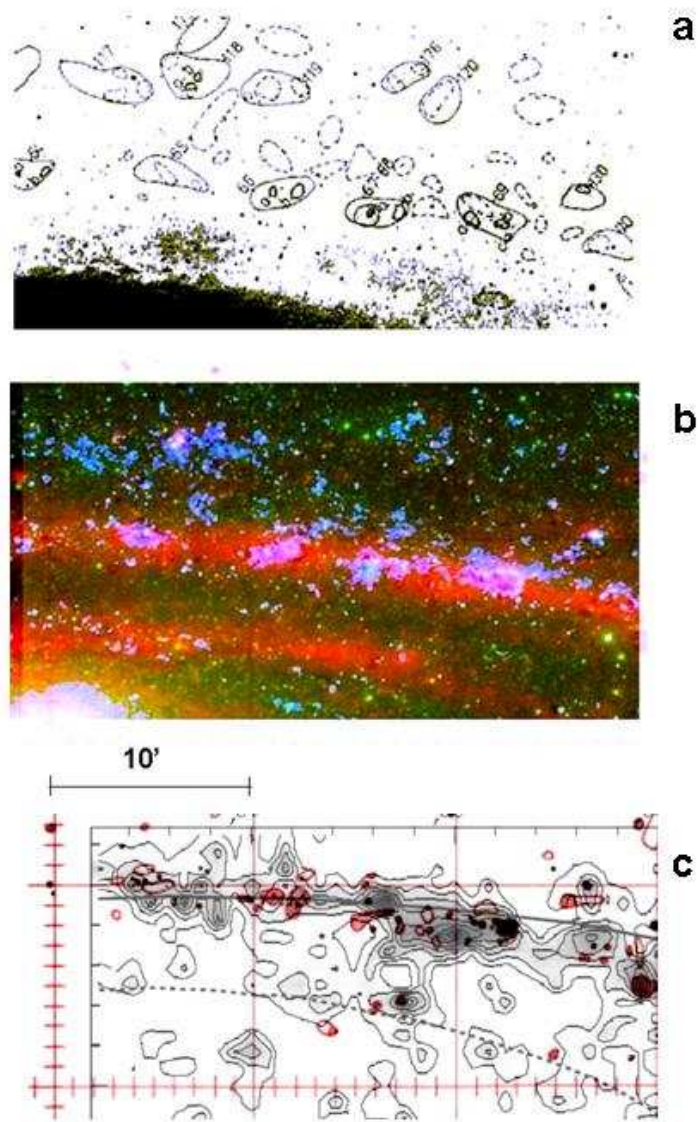

**Figure 3.** Star complexes and CO gas in the segment of the northwestern arm:
(a) Star complexes outlined by eyes from UBV plates obtained with 2-m Rozhen reflector
 (Efremov et al. 1987).
(b) Enhanced version of the GALEX and Spitzer telescope overlay image.
 (c) HII regions (Pellet et al. 1978, the most bright regions are black ones) and CO gas (Loinard
et al. 1999). Contrary to HII regions, CO clouds do not correspond well to star complexes,
which might be due to O-stars/HII regions actions.

## 3 Complexes in the northwestern arm and magnetic field

Rather regular spacing of complexes is seen in Fig. 2 over all the northern arm, but
it is the most regular and accompanied by the same size of complexes just in one segment of
this arm (Figs. 2 and 3). We revealed that the greatest regularity in the distribution of
complexes along the arm is seen precisely in the northwestern arm segment where, as was
found by Beck et al. (1989), the maximum degree of regular (along the arm) polarization is
observed at wavelengths of 20.1 and 6.3 cm (Fig. 4). In this arm segment, Beck et al. (1989)

found a wavy variation in the degree of polarization at 6.3 cm for the range of galactocentric distances 8–10 kpc, with the wavelength being 2.3 kpc. As we see from Fig.4, it is just in this distance range 8–10 kpc that the arm segment where we noticed a regular distribution of complexes is located.

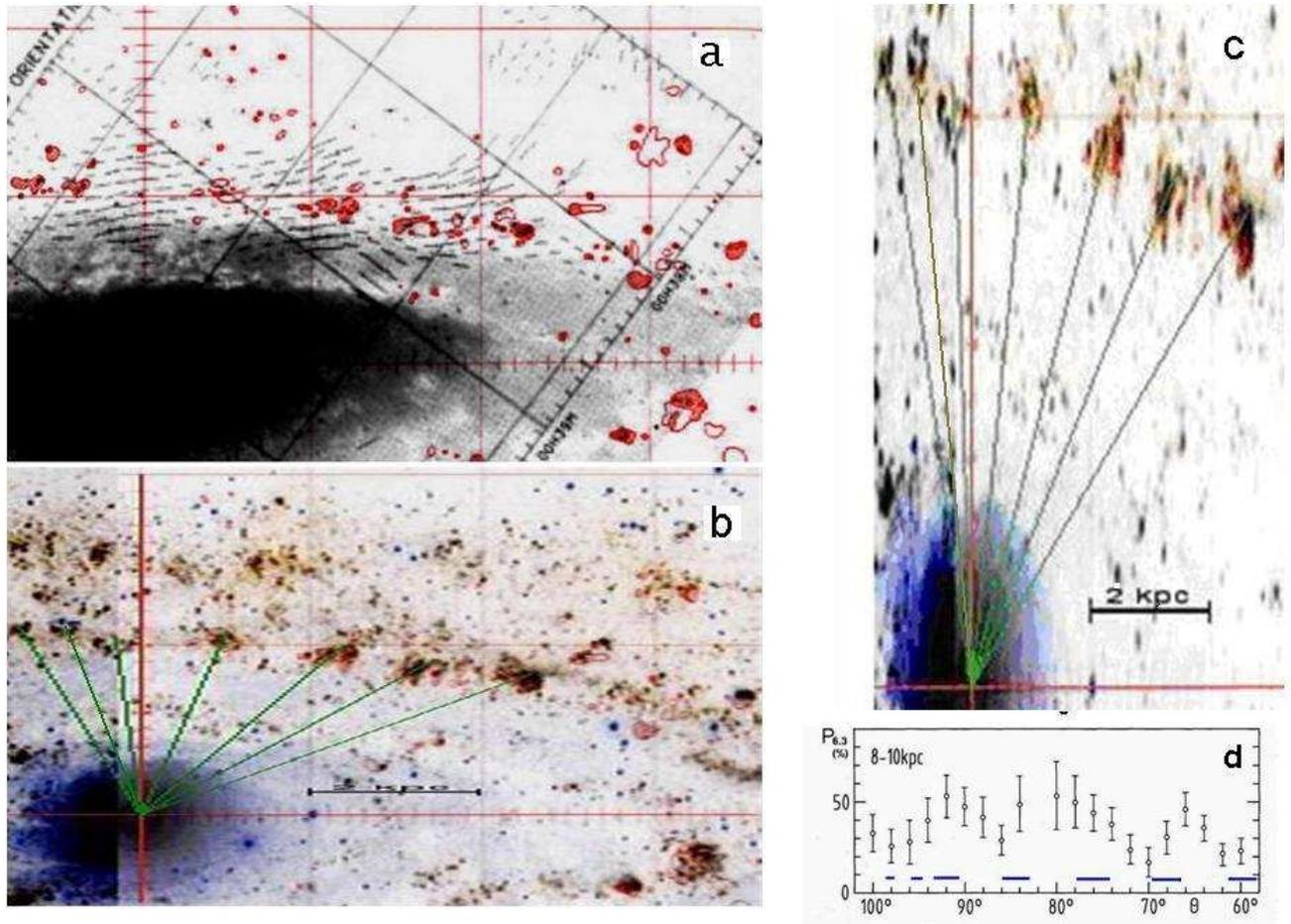

**Figure 4.** Star complexes and magnetic field in the segment of the northwestern arm.
(a) The superposition of Fig. 4 from Beck et al. (1989), which presents the polarization vectors at λ6.3 cm, on the HII image of the NW part of M31
(b) The GALEX image of the same part of M31, the azimuthal angles of complexes are shown.
(c) The image of the star complexes in the segment of NW arm reduced to on-pole position.
The M 31 plane inclination is assumed to be 12 degree, the same as used by Beck et al. (1989). The lines drawn from the center of M31 point to the centers of the star complexes, they are the same as in Fig. 5b.
(d) Degree of polarization versus azimuthal angle at the center of M31 ((from Fig. 8 from Beck et al. 1989) and the azimuthal angles of star complexes (blue stretches). . The azimuthal angle in the M31 plane (0° along the major axis, 90° along the minor NW axis) is given along the horizontal axis; the degree of polarization at λ6.3 cm between the distances of 8 and 10 kpc from the center of M31 is along the vertical axis; the azimuthal angles of star complexes (obtained from Fig. 4c) are marked at the bottom of the figure.

The polarization wavelength in this segment is approximately twice the derived average spacing between five star complexes in this region. The average size (along the arm) of five complexes located between azimuthal angles 94 − 61° (see Fig. 4d) is 0.60 kpc and their mutual average distance along the arm is 1.2 kpc. (The smaller groups at PA 98 and 95 might be considered as the part of one more complex)

The spacings between the complexes along this arm segment roughly correspond to those between the maximum and minimum degrees of polarization: the complexes clearly have a tendency to be located near the polarization extremes (Fig. 4d). Beck et al. (1989) concluded that the relationship of the polarization parameters to the magnetic field orientation and their variation along the arm are indicative of a three-dimensional wavy magnetic field structure probably related to the Parker–Jeans instability. We may now say that the formation of star complexes along the arm at regular distances that are approximately half the wavelength of this instability is also related to the latter.

It looks like the wavy structure of the gas/dust lane, seen in Figs. 2b and 2c, may be just an optical evidence of the gas lane shaping under action of the Parker-Jeans instability. Most star complexes are located within this lane in wavy-like way as well – they deviate from the average line of the dust lane. The wavy appearance of the lane and complexes location there most probably explained by their deviations from the M31 average plan, which seems to be natural within framework of magneto-gravitational instability scenario (Franco et. al. 2002, Lee et al. 2004, Mouschovias et al. 2009). If the wave extremes lie at extreme distances from the M31 plane, then the distances of their centers from the M31 plane can be estimated to be about 50–100 pc, assuming the inclination of the galaxy plane to the line of sight is $12°$. In our Galaxy both complexes of young clusters studied by Alfaro et al. (1992) in the Carina – Sagittarius arm deviate by ~30–40 pc from the Galactic plane.

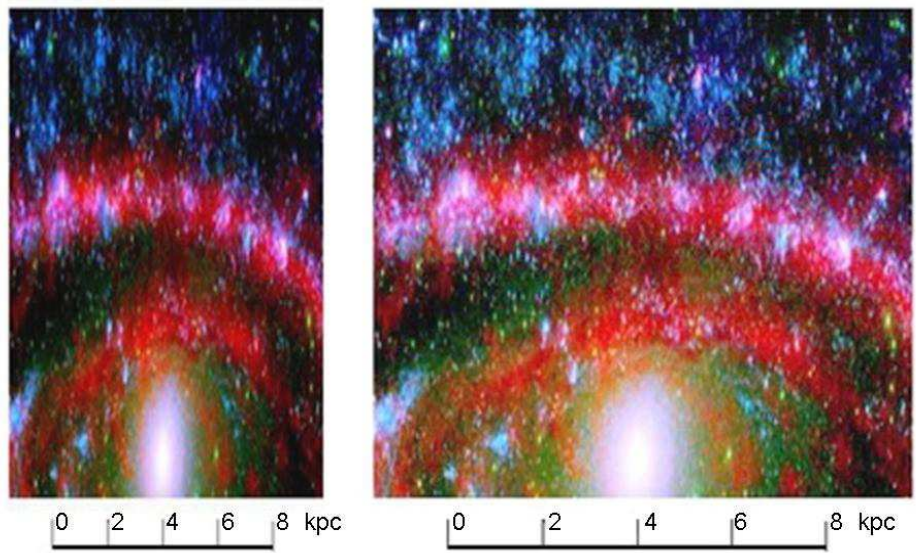

**Figure 5.** On-pole images of the northwestern arm.
Left-hand panel: inclination of the M31 plan is $12°$.
Right-hand panel: inclination of the M31 plan is $16°$.

In the pole-on image of M31, reduced with inclination $12°$ (adopted by Beck et al. 1989) the complexes look elongated in direction across the arm (Figs. 4c and 5), but they transform to round ones after assuming inclination $16°$. The latter value is just found recently from HI data for the distances from M31 center, corresponding to this arm segment position (Chemin et al. 2009). The M31 plan is clearly non-planar.

It looks like the regular complexes in an arm of a grand design galaxy (presumably being bound and then round ones) might give the quite new way to determine (better say to check) the local value of a galaxy plan inclination to the line of sight. Another possibility to explain their peculiar visible shape (under inclination 12°) is to assume (within the northwestern arm) the respective inclination of every complex plan to the galaxy plan. Such an inclination is known for the Gould Belt complex and some others in the Milky Way. Note that nor sizes neither spacing of complexes along this arm segment (roughly parallel to the major axis) are affected by the small difference in the accepted inclination (Fig. 5).

## 4 Structure of the southwestern arm

In Figs. 6 and 7 a sharp difference in the structures of the northwestern and southwestern arms (parts of the same Baade's S4 arm) in M31 is seen. In the former, the complexes are located inside the gas–dust arm; in the latter, individual complexes cannot be identified at all, while the thick gas/dust lane lies inside (upstream) the bright stellar arm. In complete agreement with the expectations of the theory of spiral arms as density waves, the H II regions here are located at the boundary of the lane of neutral gas and stars (Fig. 6) and a gradient in stellar ages is observed across the arm (Efremov 1985, 1989).

The possible explanation for this correspondence to the expectations of the density wave theory of spiral arms is as following. We see from Fig. 7 that the bright blue arm segment (densely populated by early-type high-luminosity stars) has an unusually large for M31 arms pitch angle $i$, about 25-30°. We also see that for the northwestern arm and, in general, for almost all of the other arms, this angle is close to zero. According to the classical theory (Roberts et al. 1975), the degree of gas compression by a spiral shock wave is determined by the component ($W$) of difference between the velocities of solid-body rotation of the density wave ($V$dw) and of differential rotation of the galaxy's gas ($V$) around its center, which is directed perpendicularly to the wave front, i.e., toward the inner boundary of the arm:

$$W = (V - V\text{dw}) \sin i.$$

The higher the density of the initial gas cloud, the more efficient the star formation. The large pitch angle of the arm segment in question leads to a high value of $W$ and, hence, to strong gas compression and determines the high star formation rate in the shock. This should lead to the easier observable transverse age gradient - younger stars are closer to their birthplaces in the gas–dust lane. Such a gradient is observed in the southwestern segment of S4 arm indeed (Efremov 1985, 1989, and Fig. 8). The presence of a shock in this segment of the spiral arm is also confirmed by its rectilinearity (Efremov 2001), which is immediately apparent in Fig. 7. According to Chernin et al. (2001), the presence of rectilinear segments in the spiral arms of galaxies is explained by the tendency for the shock wave to straighten out its front.

Anyway, the segments of wave spiral arms of such a kind should be distinguished from the transient kneed spiral arms that result from the gravitational instability of galactic disks. Such arms have long straight segments, but their transverse profile should be symmetric about the gravitational potential well ("gorge") and their stratification into gas and stars should be absent (Dobbs and Bonnell 2008).

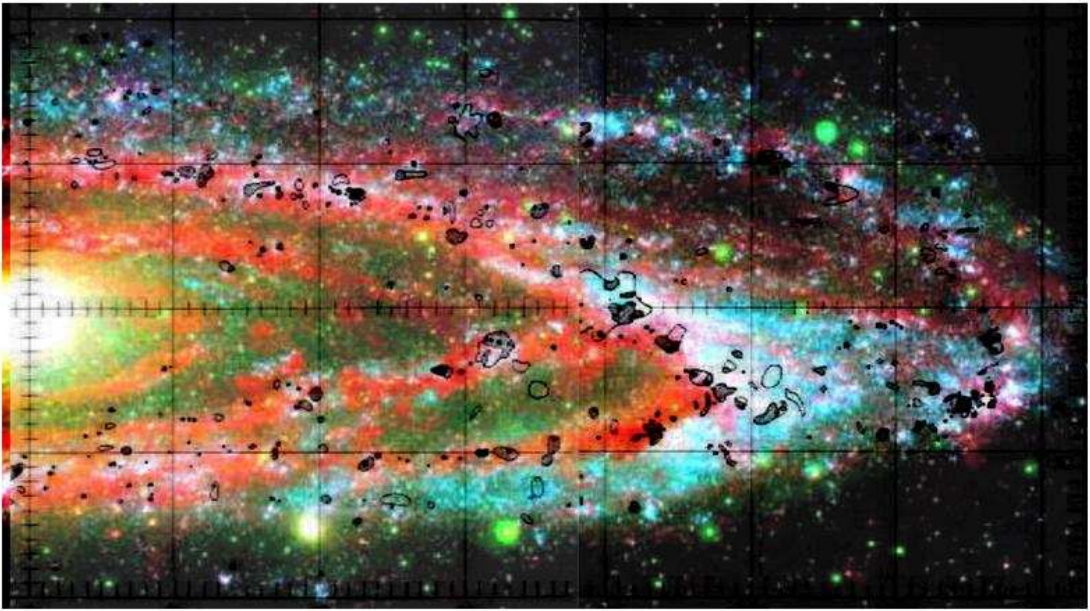

**Figure 6.** GALEX + Spitzer image overlaid onto HII regions (black) map from Pellet et al. (1978).

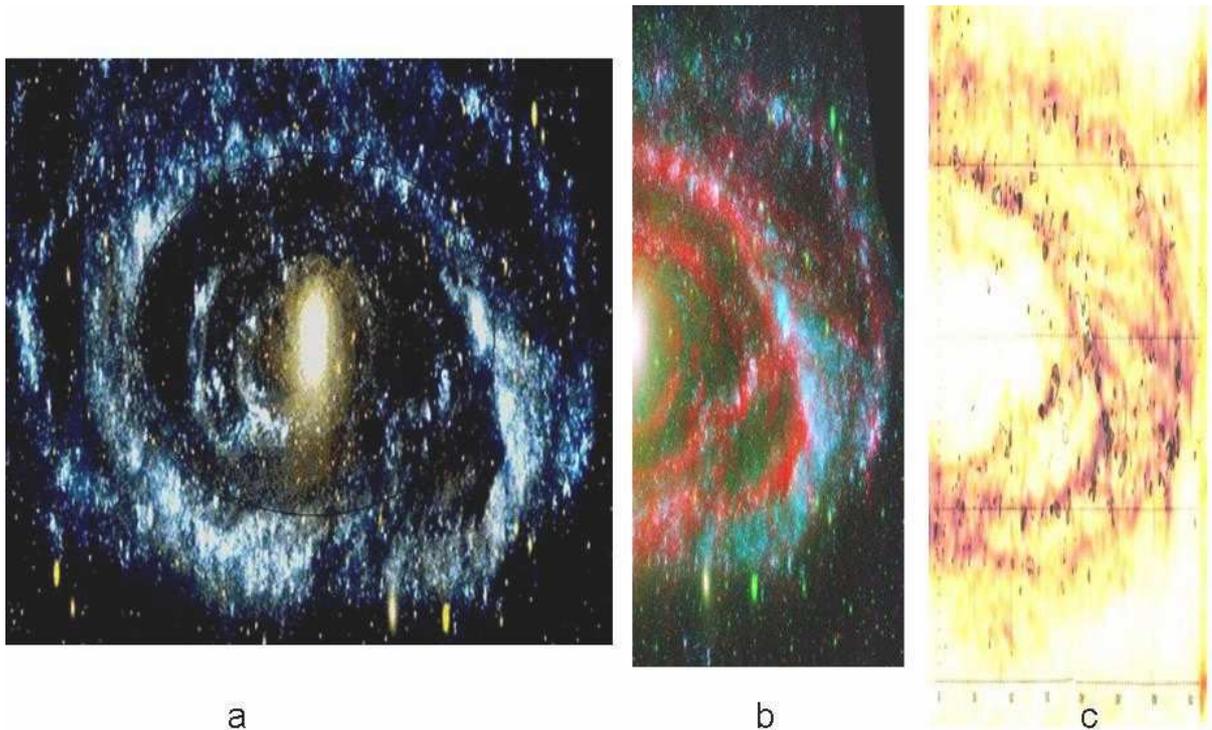

**Figure 7.** Pole on-images of M31, assuming the plane inclination $12°$. NW is at the top, SW is to the right-hand side.
(a) GALEX + circle through NW and SW arm segments.
(b) GALEX + Spitzer telescope image.
(c) HI (Braun et al. 2009) + HII (Pellet et al. 1978).

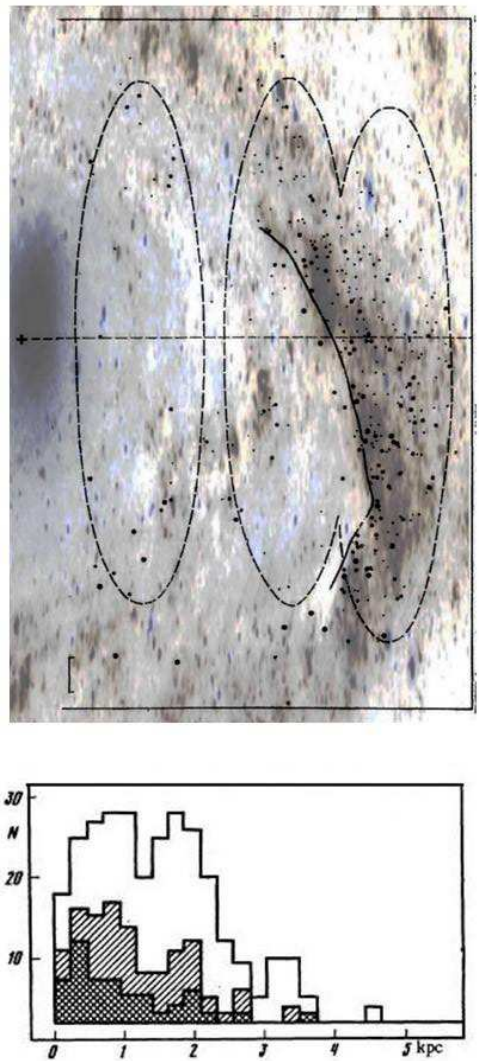

**Figure 8**. Cepheids in the Andromeda galaxy.
Top panel: Baade's field I, II and III where Cepheids (black points) were investigated (figure from Efremov 1989) overlaid onto GALEX image. The inner edge of the S4 arm is shown. The vertical stretch in low left corner is equal to 1 kpc.
Bottom panel: distribution of Cepheids in the distance from the S4 arm edge (Efremov 1989). Upper histogram – all the periods, median – periods longer then 10 days, bottom histogram - periods longer then 15 days. According the period-age relation, the longer period, the younger a Cepheid ( Efremov 2003). Note that these ages do not suffer from the light extinction.

## 5 Spiral arm structure in the Grand Design galaxies

We argued that it is the magneto-gravitational instability which determines the same spacing between complexes and their similar sizes observed in one of the M31 arm. The pure gravitational instability along an arm is surely able to do the same. In the (presumably tidal) arm connecting the interacting galaxies NGC 2207 and IC 2163, D.Elmegreen et al. (2006) found the sizes of the complexes to be distributed according the normal law - as plausibly in tidal arms in general, and as the recent HST image of Stephan Quintet demonstrated too.

In  all these cases, there is definitely no a spiral shock wave. A spiral shock can probably prevent the fragmentation of the arm segment into superclouds (parental for star complexes), because it rapidly leads to a high gas density in the entire segment and, as a result, to star formation everywhere in it. Any process leading to fast formation of numerous small scale sites of star formation in an arm may probably prevents the formation of superclouds in the corresponding arm segment.    In contrast, if there is no fast  ubiquitous star formation, the conditions for the development of a large-scale instability are retained along the gaseous arm and it fragments into superclouds, - only within which a gas density sufficient for star formation is initially reached – in such a way the star complexes may emerge.

The low rate of star formation is probably needed   also to  formation of the regular magnetic field along an arm.  Important conclusions  can be drawn from data on the magnetic field  in NGC 6946. Beck (1991) found that the regular magnetic field in this galaxy forms magnetic arms that are located between the gaseous–stellar arms and that have the same pitch angle, with the degree  of field uniformity anticorrelating with the intensity of neutral and ionized gas lines. He explained this  as the result of a high star formation rate in the stellar–gaseous arms of NGC 6946, which causes the magnetic field lines in them to be tangled. Recently, Chyzy (2008) offered a similar explanation for the anticorrelation between the star formation rate and the degree of magnetic field regularity that he found in NGC 4254. Note that the star formation rate in M31 is low.

Note also  the impossility  to divide the spiral arms of NGC 6946 into separate star complexes; they are entirely filled with sites of  star formation and young stars.  The only (and quite peculiar,  Efremov et al. 2007) bona fide complex  in this galaxy is  outside spiral arms.

Considering that the obligate  signature  of the spiral shock wave is the dark lane upstream of a stellar arm,  we  may expect that  the  regular chain of complexes should  be located along the arm segments where such  a lane is absent.    This conclusion agrees with the results of examining the images of several galaxies containing chains of complexes.  For example, in M74  no individual star complexes can be identified in the shorter arm, with the dust lane running in front of its inner edge, while there are seven widely separated complexes in the other, longer arm, with no dust lane running upstream  this arm edge. However,  in the middle of the latter arm there is a dust lane,  connecting these complexes, as can be seen in the infrared images of the Spitzer telescope,  recording warm dust  (Fig. 9).

B.Elemegreen et al. (2006) studied the   distrubution in sizes for young stellar groupings  in M74 but  studied region was limited by the galaxy central region. They  found the cumulative size distribution to be a power law in all passbands, with a slope of approximately  -1.5  over  1.8  orders  of  magnitude. The  luminosity  distribution  is approximately a power law as well, with a slope of approximately -1 for logarithmic intervals of luminosity. Their conclusion was that these  results suggest a scale-free nature for stellar aggregates in a galaxy  (and therefore no star complexes are there).  However  the studied region  included the shorter arm, but not encompassing the longer arm with its about equally spaced star complexes.  The latter region should give excess of the large sizes and regular spacing,  found  in this galaxy by Elmegreen and Elmegreen (1983) long ago.

a               b               c

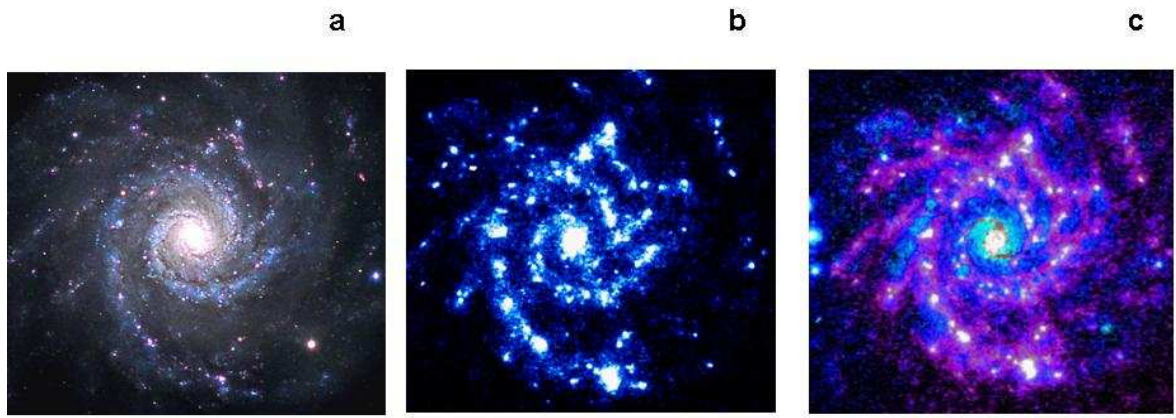

**Figure 9.** M 74 (NGC 628) galaxy. North is up.

(a) An optical image. No dust lane is seen upstream of the SE (longer) arm, but this lane is upstream another arm, which hosting no complexes.

(b) UIT image. Note the regular chain of large complexes along the SE arm.

(c) Overlay of UIT image on the Spitzer telescope far-IR image. The lane of the warm dust (and therefore gas) connected the star complexes is seen now, like it is the case in the NW arm of M31.

The images of M51 galaxy seemingly confirm the observation that star complexes and dust lane are mutually excluded. Note the OB-associations at North of Fig. 10b – they are spaced in quazi-regular way and located down-stream the dark lane (Bastian et al. 2005) but they are too small and compact for to be called complexes. The latters are much larger and a few of them are located in southern part of the same arm, where there is no dust lane (Fig. 10).

Rather similar is the situation in NGC 2997, where Grosbol and Dottori (2009) found many associations (HII regions), located immediately behind the dust lane, and the age gradient across the arm. Anyway a few of the large complexes are seen in the outer distant part of the arm. Note that the difference in the velocities of the density wave and a galaxy stuff depends not only on the pitch-angle, but first of all on the distance from the corotation (where this difference is zero) and it surely explains why the complexes are mostly located in the distant part of arms.

. It worth noting that for the distant galaxies and/or in images obtained under the poor resolution the sizes of bright 0B-associations (HII regions) appear quite large and they may be wrongly identified as complexes - located downstream the dark lane.

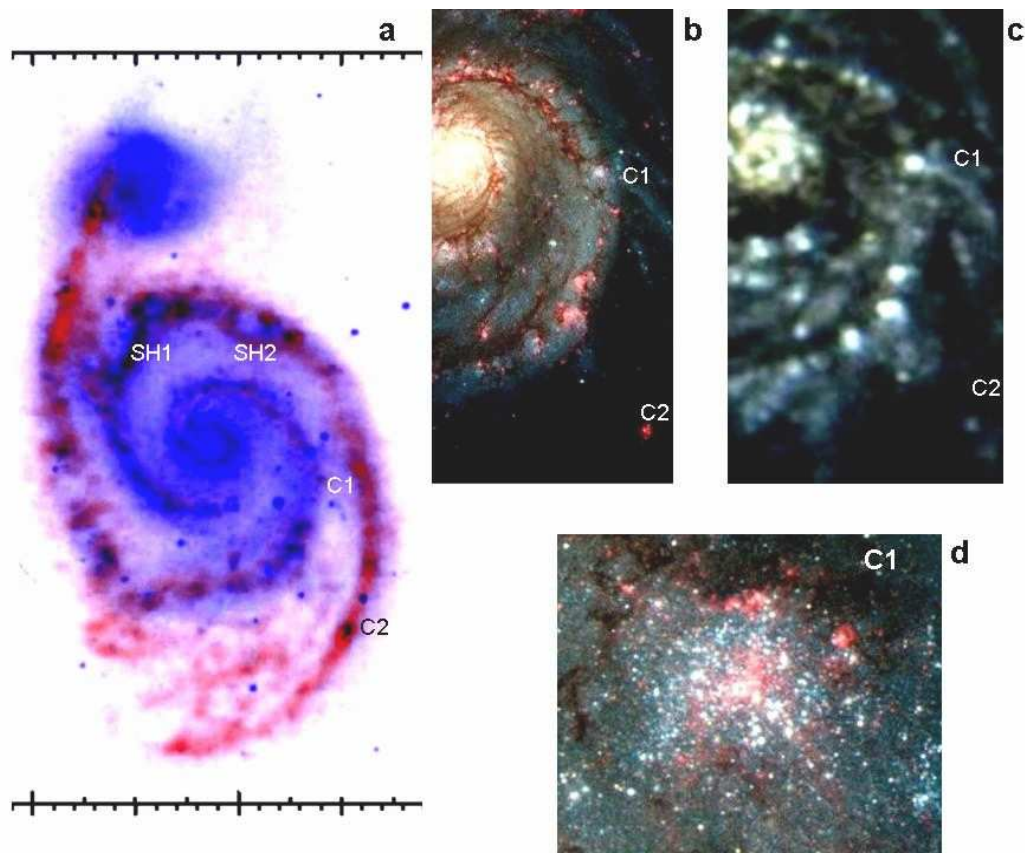

**Figure 10.** M 51 (NGC 5194) galaxy and details of its structure.
(a) Overlay of B (blue) and HI (red) images (based on THINGS' data). Two unstudied supergiant superholes (SH1 and SH2) are noted, both being located within the spiral arm. Small young complex marked as C2 located inside the outer HI/dust arm (which is of amazingly perfect circular shape), it is bright in Ha. Similar complexes are also in the arm connected NGC 5194 and NGC 5195.
(b) The western part of M 51, showing the spiral arm with chain of associations downstream the dark lane in the northern part of the arm, and four large complexes located to South, in the arm segment hosting no dark lane.
(c) The same part of M 51 galaxy, imaged by GALEX. Only four complexes are there, complex C1 hosting about no HII regions. Note the long spur started from this complex (G1 in Bastian et al., 2005). Unusual abundance of the star clusters is seen in this spur in the HST ACS image.
(d) The best example of what should be called "a star complex", with size about 400 pc, a few small HII regions and notable substructure. It is nor an OB-association, neither a giant HII region ("superassosiation"). This complex is designated as G1 in Bastian et al. (2005).

# 6 Conclusions

We have found that in M31 the regular magnetic field is observed precisely in the arm segment (Beck et al. 1989) that contains the most regular chain of star complexes with the spacing between them equal to half the wavelength in the magnetic field structure.
It may be that regular chains of complexes emerge always in spiral arms if they have a regular magnetic field, the condition for which is a rather low star formation rate. The operating formation mechanism of such chains is the Parker–Jeans instability, as was suggested in the first study where such chains were discovered (Elmegreen and Elmegreen 1983, Elmegreen 1994). However, the pure gravitational instability is surely able to do the same job as well, as frequent presence of chain of complexes along the tidal arms demonstrates.

Star complexes are not the ubiquitous units for star formation, though are the largest ones. The existence of a preferential, physically determined average size distinguishes the star complexes (at least those located within the grand design spiral arms) from all of the other classes of stellar groups. If so, to check the presence of complexes, we should construct the size distribution function for young and moderately young (i.e., all those visible in the GALEX images) discrete stellar groups in a given galaxy—this presence will manifest itself as an excess (compared to the distribution described by an power law) of clumps with large sizes. The sizes of proper complexes should be distributed according the normal law, as D.Elmegreen et al. ( 2006) have found in the arm connecting NGC 2207 and IC 2163.

M31 hosts many dozen of such groupings, whereas M51 some six or seven only, and M81 - none. These galaxies confirm the proposition that a spiral shock wave is an obstacle to a complex formation.

Apart from the tidal arms, it may be that regular chains of star complexes occur only when magnetic field in the arm is regular and that requires the star formation rate to be modest. It looks like there is no spiral shock wave in the parts of the arms where the chains of complexes are observed. These suggestions are needed to be checked with more observational data.

## Acknowledgements

I am grateful to B.G.Elmegreen for many useful discussions and to A.V.Zasov for some comments. The possibility to use the images available at GALEX and Spitzer space telescopes sites is greatly appreciated. This work was supported by the grant for support of leading scientific schools (NSh-433.2008.2).

## References

Alfaro E., Cabrera-Cano J., Delgado A.,1992, ApJ, 399, 576
Baade W., 1963, Evolution of stars and galaxies, Harvard U. Press
Bastian, N., Gieles, M.; Efremov, Yu. N.; Lamers, H. J. G. L., 2005, A&A, 443, 79
Beck R., 1991, A&A, 251, 15
Beck R., Loiseau N., Hummel E., Berkhuijsen E. M.; Grave,R.; Wielebinski, R., 1989,
    A&A, 222, 58
Bergh van den S., 1964, ApJS, 9, 65
Braun, R.; Thilker, D. A.; Walterbos, R. A. M.; Corbelli, E., 2009, ApJ, 695, 937
Chemin L., Carignan C., Foster T., 2009, arXiv: 0909.3846v1
Chernin A.D., Kravtsova A.S., Zasov A.V., Arkhipova V.P., 2001, Astron. Rep. 45, 841
Chyzy K.T.,2008, A&A, 482, 755
Dobbs C.L., Bonnell I.A., 2008, MNRAS, 385, 1893
Efremov Yu.N., 1979, Sov.Astron. Lett. 5, 12
Efremov Yu.N., 1985, Sov. Astron. Lett. 11, 69
Efremov Yu.N., 1989, *Sites of Star Formation in Galaxies: Star Complexes and Spiral
    Arms,* (Fizmatlit, Moscow, 1989), ch. 8 [in Russian].
Efremov Yu.N., 1995, AJ., 110, 2757
Efremov Yu.N., 1998, Astron. Astrophys. Trans. 15, 3
Efremov Yu.N., 2001, Astron. Astrophys. Trans. 20, 115
Efremov Yu.N., 2003, Astron. Rep, 47, 1000
Efremov Yu.N., 2004, Astrophysics, 47, 273
Efremov Yu.N., 2009, Astron. Lett., 35, 507

Efremov Yu.N., G. R. Ivanov G.R.,  Nikolov N.S., 1987, Astrophys. Space Sci. 135, 119

Efremov Yu. N.; Elmegreen B. G.,  1998,  MNRAS, 299, 588

Efremov Yu.N., Afanasiev V.L., Alfaro  E.J. et al., 2007,  MNRAS, 382, 481

Elmegreen B.G., 1994, ApJ, 433, 39

Elmegreen B.G., 2009,  IAU Symp. 254, 289

Elmegreen B.G., Efremov Yu.N., 1996, ApJ, 466, ..802

Elmegreen B.G., Elmegreen  D.M., 1983, MNRAS, 203, 31

Elmegreen B.G.; Elmegreen D.M., Chandar R.; Whitmore B. Regan M., 2006, ApJ, 644,.879

Elmegreen D.M., Elmegreen B.G., Kaufman M.  et al., 2006,  AhJ, 642, 58

Franco J., Kim J., Alfaro E.J., Hong S.S., 2002, ApJ, 570, 647

Grabelsky D.A.,  Cohen R.S., Bronfman L., Thaddeus P., 1987, ApJ, 315, 122

Grabelsky D.A.,  Cohen R.S., Bronfman L., Thaddeus P., 1988, ApJ, 331, 181

Grosbøl P.; Dottori H.,   2009, A&A, .499L, 21

Lee  S.M.,  Kim J., Franco J., Hong S.S., 2004, J. Korean Astron. Soc. 37, 249

Loinard, L.; Dame, T. M.; Heyer, M. H.; Lequeux, J.; Thaddeus, P., 1999, A&A, 351, 1087

McGee R.X., Milton J.A., 1964, Austral. J. Phys., 17, 128

Mouschovias T.Ch., Kunz M.W., Christie D.A., 2009,  arXiv:0901.0914v1 [astro-ph]

Neininger, N.; Nieten, C.; Guelin, M.; Ungerechts, H.; Lucas, R.; Muller, S.; Wielebinski, R,.
2001,  IAU Symp.#205, ed. R. T. Schilizzi,  p. 352

Odekon M.C., 2008, ApJ, 681, 1248

Pellet A., Astier N., .Viale A. et al., 1978, A&A Suppl. Ser. 31, 439

Roberts W.W., Roberts M.S., Shu  F.H., 1975, ApJ, 196, 381